\documentclass[12pt]{iopart}
\usepackage[english]{babel}
\usepackage{amssymb}
\usepackage{graphicx}

\begin{document}

\title{Survival probability of an immobile target surrounded by mobile traps}
\author{Jasper Franke $^{(1)}$ and Satya N. Majumdar $^{(2)}$ }
\address{$^{(1)}$ Institut f\"ur Theoretische Physik, Universit\"at zu
K\"oln, K\"oln, Germany\\
$^{(2)}$ Univ. Paris-Sud, CNRS, LPTMS, UMR 8626, Orsay F-01405, France}

\noindent 
\begin{abstract}
We study analytically, in one dimension, the survival 
probability $P_{s}(t)$
up to time $t$ of an immobile target surrounded by mutually noninteracting
traps each performing a continuous-time random walk (CTRW) in continuous space.
We consider a general CTRW with symmetric and continuous (but otherwise
arbitrary) jump length distribution $f(\eta)$ and arbitrary waiting time 
distribution
$\psi(\tau)$. The traps are initially distributed uniformly in space with
density $\rho$.
We prove an exact 
relation, valid for
all time $t$, between $P_s(t)$ and the expected maximum $E[M(t)]$ of the
trap process up to time $t$, for rather general stochastic motion $x_{\rm
trap}(t)$ of each trap. When $x_{\rm trap}(t)$ represents a general CTRW with
arbitrary $f(\eta)$ and $\psi(\tau)$, we are able to compute exactly
the first two leading terms in the asymptotic behavior of $E[M(t)]$ for
large $t$. This allows us subsequently to compute the precise asymptotic
behavior, $P_s(t)\sim a\, \exp[-b\, t^{\theta}]$, for large $t$, with exact
expressions for the stretching exponent $\theta$ and the constants $a$ and
$b$ for arbitrary CTRW. By choosing appropriate $f(\eta)$ and $\psi(\tau)$,
we recover the previously known results for
diffusive and subdiffusive traps. However, our result is more general and
includes, in particular, the superdiffusive traps as well as totally
anomalous traps.

\end{abstract}

\maketitle
\section{Introduction}
The study of the survival probability of a target particle surrounded by 
diffusing traps 
is a classic problem in physical chemistry that goes back to Smoluchowski in 
1916 \cite{Smolu}. 
Smoluchowski introduced this problem to
compute the rate of chemical reactions in the diffusion-driven regime. 
The basic model and its variants~\cite{Chandra,Redner} have since found a large 
number of 
applications
in a wide variety of contexts such as in reaction-diffusion systems~\cite{Rice},
chemical kinetics~\cite{Benson,OZ,TW,Ziff1,MCZ,ZMC1,ZMC2},
predator-prey models in  
population dynamics~\cite{RK,Redner} and also in a wide class of the
so called `walker persistence'
problems~\cite{WP}.

In the simplest setting of the model known as the `target annihilation' 
problem~\cite{TA}, 
the target particle is immobile at the origin and the {\em noninteracting} 
traps, initially distributed uniformly over space with a finite
density $\rho$, each  
perform an unbiased continuous-time Brownian motion with a diffusion
constant $D$.  
When any of these
traps hits the target particle at the origin, the target particle gets
annihilated.   
The average survival probability $P_s(t)$ of the target up to time $t$,
where the average is over the initial positions of the traps distributed
uniformly over space with density $\rho$, 
has been 
computed
exactly~\cite{TA} in all dimensions.
In particular, in the simplest one dimensional case, $P_s(t)$ decays with $t$
as a stretched exponential for all $t$,
\begin{equation}
\label{target}
P_{s}(t)=\exp\left[-4\rho\sqrt{\frac{Dt}{\pi}}\right].
\end{equation}
In $d=2$, $P_s(t)\sim \exp[-\rho c_2 t/\ln t]$ while
for $d>2$, $P_s(t)\sim \exp[-\rho c_d t]$ for large $t$, where
the constants $c_2$ and $c_d$ are known~\cite{TA}. 
In $d> 2$, the target particle needs to have a finite size while
for $d\le 2$, the target can be considered to be a point particle.
In the original Smoluchowski version of the problem, the quantity
$F(t)= -\ln [P_s(t)]$ has a physical interpretation: it represents
the net flux of traps to a static absorber at the origin up to time 
$t$~\cite{Ziff1,MCZ,ZMC1}.
In $d=1$, the net flux $F(t)=4\rho\sqrt{\frac{Dt}{\pi}}$ thus grows 
algebraically $\sim t^{1/2}$ with time $t$.

The original Smoluchowski or equivalently 
the `target annihilation' problem has been generalized in the literature in 
several ways, e.g., 
(a) when the target itself moves stochastically (not necessarily
diffusive) while  
the noninteracting
traps perform normal diffusion
and (b) when the target is static but the traps, while still noninteracting, 
perform, in general, non-diffusive stochastic motion (c) when both the
target and  
the traps
move nondiffusively in general and (d) the famous Donsker-Varadhan 
problem~\cite{DV}
where the target moves diffusively but the traps are immobile. In this
paper, we  
will focus on the cases (a) and (b) only.

An interesting variant belonging to class (a) above is the 
`diffusive target 
annihilation'
problem where the target particle itself diffuses with a diffusion constant 
$D_0$~\cite{BL}. This problem has seen a recent flurry of 
activity~\cite{MG,BB,OBCM,BMB,ABB} and it has been proved rigorously
that in one dimension, the average survival probability $P_s(t)$ of
the diffusing  target
particle has the same asymptotic stretched exponential decay as in
Eq. (\ref{target}) 
(and hence is independent of $D_0$), though with considerable subleading 
corrections to the leading behavior for intermediate $t$. Another 
variant in class (a) is the `ballistic target annihilation' problem where the
target particle moves ballistically with a constant velocity $c$. In this
case, the survival probability decays faster, $P_s(t)\sim \exp[-\vartheta t]$
in all dimensions $d$, where the inverse decay rate $\vartheta$ has
been computed exactly for $d\le 2$ and $d=3$~\cite{MB}. 

The variants in class (b) where the target is static and the traps 
undergo subdiffusive motion or (c) where both the target and the traps
undergo subdiffusive motion have been 
studied extensively by Yuste and collaborators~\cite{YL,YRL,YL1,YOLBK,BAY}.

Another recent extension belonging to class (b) above is to the case
where the target is static, each trap
performs independent Brownian motion but resets to its own initial position
with a constant rate $r$~\cite{Reset}. This problem has interesting
implication in the context of search problems where a team of searchers (traps)
adopt the resetting strategy to make the search of the target more 
efficient~\cite{Reset}.  
For $r=0$, it reduces to the standard 
target annihilation 
problem. For a nonzero $r$, the average survival probability in one dimension
was computed exactly and it was found to decay 
as a power-law at late times~\cite{Reset}
\begin{equation}
P_s(t)\approx C\, t^{-2\rho\sqrt{D/r}}
\label{reset1}
\end{equation}
where $C$ is a constant. For short times $rt<<1$, the decay is 
stretched-exponential~\cite{Reset} as in Eq. (\ref{target}).

In this paper, we study analytically the average survival probability
of a target 
particle in one dimension in another variant belonging to class (b). 
In this model, the target 
remains immobile at the origin as in the original target annihilation 
model. The traps also remain noninteracting and are distributed initially 
in space with a uniform density $\rho$.  
However, each trap,
instead of performing a continuous-time Brownian motion with
a diffusion constant $D$, now performs a continuous-time random walk (CTRW)
with arbitrary jump length distribution and arbitrary waiting time distribution 
between successive jumps.
Our result for the asymptotic survival probability includes
the cases when the traps undergo subdiffusive as well as superdiffusive motion,
thus generalizing the previously known result for subdiffusive traps~\cite{YL}.

In the standard version of CTRW, a trap moves by making
successive random jumps at random times~\cite{MS,BG,MK}. 
More precisely, starting from its initial position, a trap
waits a certain amount of random time $\tau$ (drawn from a normalized 
distribution $\psi(\tau)$) 
and then jumps by a certain distance $\eta$ (also a random variable
drawn from a continuous and symmetric distribution $f(\eta)$) (see Fig. 
(\ref{ctrw.fig})). 
At its new position
it again waits for a random
time drawn from $\psi(\tau)$ (each time independently) and then jumps
to another new position by a random distance drawn from $f(\eta)$
and the process continues. 
Thus, CTRW has two `input' distributions, one
for time and one for space, namely $\psi(\tau)$ and $f(\eta)$.
We also assume that these two distributions don't vary from trap
to trap, all trap motions are characterized by the same $\psi(\tau)$
and $f(\eta)$.
In this model, the immobile target particle at the origin gets
annihilated whenever any trap trajectory {\em goes past the origin}.
Note that the particle at the origin gets annihilated even when a trap
{\em jumps over} the origin. Thus the survival probability $P_s(t)$
of the target particle in this case is just the probability that
not any of the traps jumps over the origin up to time $t$.
Our goal is to understand how the two arbitrary input distributions
$\psi(\tau)$ and $f(\eta)$ affect the canonical 
stretched exponential decay of $P_s(t)$ 
in Eq. (\ref{target}) that is valid when each trap performs continuous-time
Brownian motion.

The advantage of CTRW is that by appropriately choosing the two
input distributions $\psi(\tau)$ and $f(\eta)$, one can study
various types of motions of a trap as special cases. For example,
if one chooses $\psi(\tau)= \delta(\tau-\Delta t)$, one generates
a discrete time jump process with jump length distribution $f(\eta)$ 
where the jumps happen instantaneously
at time steps separated by $\Delta t$. If one further
chooses a Gaussian jump length distribution $f(\eta)= \exp\left[-{\eta}^2/{2 
{\langle \eta^2\rangle}}\right]/\sqrt{2\pi {\langle \eta^2\rangle}}$
and takes the limit ${\langle \eta^2\rangle}\to 0$, $\Delta t\to 0$ but keeping
the ratio ${\langle \eta^2\rangle}/(\Delta t)=2\,D$ constant, one recovers 
the continuous-time Brownian motion. Similarly, by choosing
a power law distribution $f(\eta)\sim |\eta|^{-\nu-1}$ for large $|\eta|$ with
$0< \nu\le 2$, one can generate L\'evy flights for a trap~\cite{BG,MK,Bertoin}.
Besides, the CTRW also provides a natural framework to generate
anomalous diffusion process, namely subdiffusive and superdiffusive 
behavior that occurs in many natural processes~\cite{BG,MK}.
It would thus be interesting to know how the average survival probability
$P_s(t)$ of a target particle decays with time when each
of the traps, though still noninteracting, performs CTRW with arbitrary 
waiting time and jump length distributions, $\psi(\tau)$ and $f(\eta)$ 
respectively. 
\begin{figure}
 \includegraphics[width=0.5\textwidth]{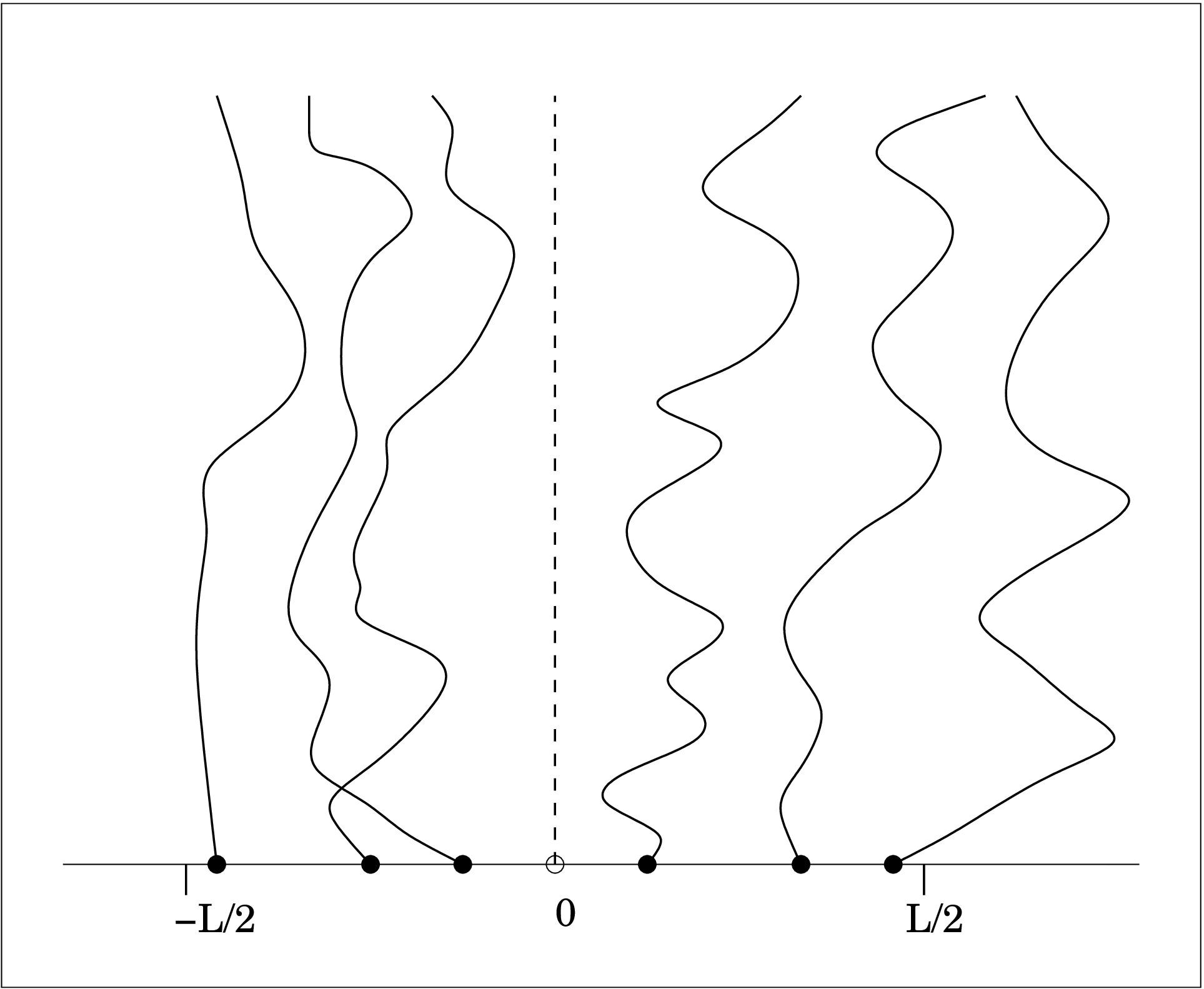}
\caption{Schematic example of a realization where the target at the origin
is immobile and there are $6$ noninteracting traps around it
undergoing stochastic  
motion. The traps are initially placed uniformly in the box $(-L/2,L/2)$.
Eventually we are interested in the limit when the number of traps
$N\to \infty$,  
the box size $L\to \infty$ but with the density of traps $\rho=N/L$ fixed.} 
\label{figure1}
\end{figure}

\vskip 0.3cm

\noindent Our main results are twofold:

\begin{enumerate}

\item We will first derive a very general result for the average survival 
probability $P_s(t)$ for any model in one dimension belonging to class
(b) above, i.e., when the target is immobile and each trap
performs independently a stochastic motion $x_i(t)$ starting from its 
initial position $x_i(0)$. The stochastic evolution law of $x_i(t)$
can be rather general. The only assumptions are (i) the process
$x_i(t)$ is invariant under a constant shift, i.e., $x_i(t)$
and $x_i(t)-c$ (where $c$ is a constant) has the same evolution equation
and (ii) the process $x_i(t)-x_i(0)$ is symmetric around the origin.
Since the evolution law of the the $i$-th trap $x_i(t)-x_i(0)$ does not depend 
explicitly on the 
index $i$, we will generally denote it by $x_{\rm trap}(t)\equiv x_i(t)-x_i(0)$.
For such a stochastic trap motion $x_{\rm trap}(t)$, we obtain
a rather general result for the average survival probability
\begin{equation}
\label{genrel}
P_s(t)= \exp\left[-2\rho E[M(t)]\right]
\end{equation} 
where $E[M(t)]$ is the expected value of the maximum of the process 
$x_{\rm trap}(\tau)$ over the time interval $\tau\in [0,t]$, i.e,
\begin{equation}
M(t) = \max_{0\le \tau\le t}\left[ \{x_{\rm trap}(\tau)\}\right].
\label{maxt1}
\end{equation}

We emphasize that this relation is very general and holds whatever be
the stochastic process that describes the motion of a trap, as long 
as the process $x_{\rm trap}(t)$ satisfies the two properties (i) and (ii)
above. 
For example, if $x_{\rm trap}(t)$ is a continuous-time Brownian 
motion, it is well known that the probability distribution of the maximum
$M(t)$ of the process is a half-Gaussian~\cite{CM,SM09}, i.e., $P(M,t)= 
\exp[-M^2/{4Dt}]/\sqrt{\pi D t}$ for $M\ge 0$. Hence the expected value
$E[M(t)]=2\sqrt{Dt/\pi}$. Plugging this result in the general relation
in Eq. (\ref{genrel}), one recovers the classical stretched-exponential
decay in Eq. (\ref{target}).

\item The relation in Eq. (\ref{genrel}) holds, in particular, when $x_{\rm 
trap}(t)$ represents
a CTRW with arbitrary $\psi(\tau)$ and $f(\eta)$. For such a general CTRW, we will 
first derive
the exact asymptotic expression of $E[M(t)]$ for large $t$. 
Plugging this result
in the general relation (\ref{genrel}) then provides us with the exact
asymptotic 
behavior of $P_s(t)$
\begin{equation}
P_s(t) \approx a\, \exp[-b \, t^{\theta}]
\label{mainres}
\end{equation}
where the stretching exponent $\theta$, the constant $b$ and the amplitude $a$
are determined exactly for arbitrary $\psi(\tau)$ and $f(\eta)$ (symmetric
and continuous) (the precise results are detailed in Eq. (\ref{surv_prob})). 
We also verify numerically our analytical predictions.

\end{enumerate}

The rest of the paper is organized as follows. In section II, we 
derive the main results of this paper: section II.A provides the
derivation of 
the general relation (\ref{genrel}) while in 
section II.B, we derive $E[M(t)]$ for CTRW by extending known
results for $E(M_n)$ in a discrete time setting \cite{CM}.
In section III, these results are put together to obtain the asymptotic
large $t$ behavior of the survival probability $P_s(t)$ for a CTRW trap
with arbitrary $\psi(\tau)$ and $f(\eta)$.
Our general result will recover, as special cases, the known
expression for traps 
performing diffusion and subdiffusion. In this section, we
also compare our analytical results to numerical simulations and obtain
good agreement. Finally we conclude in section IV with a summary and some
open questions.

\section{Derivation of the main results}

\subsection{Survival probability for general stochastic motion of a trap}
\label{Main_A}

We consider an immobile target at the origin in one dimension, surrounded 
by
traps, initially distributed uniformly over space with density $\rho$ and
each performing an independent stochastic evolution $x_i(t)$ starting
from the initial position $x_i(0)$. 
We will denote the process $x_i(t)-x_i(0)$ by $x_{\rm trap}(t)$ that
starts at the origin and satisfies two properties (i) and (ii) mentioned
in the introduction.
In order to derive an expression for
the survival probability of the target at the origin, we follow the
strategy used by Bray and Blythe~\cite{BB} for the special case when $x_{\rm 
trap}(t)$ is a Brownian motion around the origin. We consider first a
finite number $N$ of mobile 
traps starting at fixed initial positions $\{x(0)\}\equiv\{x_1(0),
\dots, x_N(0)\}$ on the interval $(-L/2, L/2)$, cf. Fig. (\ref{figure1})
and calculate the survival probability of the target up to time $t$ for this
fixed set of initial positions. Next we average this survival probability
over the starting positions where each $x_i(0)$ is drawn independently from a 
uniform 
measure on 
$(-L/2,L/2)$ and
take the limit $N\to\infty, L\to\infty$  while keeping the density
of traps $N/L=\rho$ fixed. 

Let $p\left(t|x_i(0)\right)$ denote the probability that a single
trap, starting at the 
initial position $x_i(0)$, does not cross the origin up to time $t$.
By independence of the $N$ particles, the survival probability of the
target factorizes into a product of the probabilities $p\left(t|x_i(0)\right)$ 
\begin{equation}
P_{s,N,L}(t|\{x(0)\})=\prod_{i=1}^{N}p\left(t|x_i(0)\right).
\label{surv.1}
\end{equation}
Now averaging over all starting positions with uniform distribution on
the interval $(-L/2,L/2)$ and noting that particles were only
distinguishable by their starting positions before averaging, 
\begin{eqnarray}
P_{s,N,L}(t)&=\left< 
\prod_{i=1}^{N}p\left(t|x_i(0)\right)\right>
= \left[\frac{1}{L}\int_{-L/2}^{L/2}dx\, 
p(t|x)\right]^N\nonumber \\
&=\left[1-\frac{1}{L}\int_{-L/2}^{L/2}dx\, 
\left(1-p(t|x)\right)\right]^N.
\label{surv.2}
\end{eqnarray}
Taking the limit
$L\to\infty, N\to\infty$ while keeping $N/L=\rho$ fixed and using the
symmetry $p(t|x)=p(t|-x)$ we get 
\begin{equation}
P_s(t)=\exp\left[-2\rho\int_0^\infty dx\, \left(1-p(t|x)\right)\right].
\label{surv.3}
\end{equation}

Note that, by definition,
\begin{equation}
p(t|x) = {\rm Prob.}\left[ x_i(\tau)> 0\,\, {\rm for}\,\, {\rm all}\,\, 0\le \tau\le 
t\,\Big|\,x_i(0)=x\right]
\label{trap.1}
\end{equation}
where $x_i(\tau)$ represents the stochastic trajectory of a trap, say the $i$-th trap, starting
at the initial position $x_i(0)=x$. Consider now the stochastic process
$x_{\rm trap}(\tau)= x_i(\tau)-x_i(0)=x_i(\tau)-x$. Then the
probability $p(t|x)$ in Eq. (\ref{trap.1}) translates into the following
probability for the process $x_{\rm trap}(\tau)$
\begin{equation}
p(t|x)={\rm Prob.}\left[ x_{\rm trap}(\tau)> -x 
\,\, {\rm for}\,\, {\rm all}\,\, 0\le \tau\le 
t\,\Big|\,x_{\rm 
trap}(0)=0\right]
\label{trap.2}
\end{equation}
Using the spatial inversion symmetry of the process $x_{\rm
  trap}(\tau)$ around the origin, Eq. (\ref{trap.2}) can be written as
\begin{equation}
p(t|x)={\rm Prob.}\left[ x_{\rm trap}(\tau)<x\,\, {\rm for}\,\, {\rm all}\,\, 0\le 
\tau\le
t\,\Big|\,x_{\rm    
trap}(0)=0\right]
\label{trap.3}
\end{equation}
which, incidentally, is precisely the cumulative probability that the maximum
$M(t)=\max_{0\le \tau\le t}[\{x_{\rm trap}(\tau)\}]$ of the process $x_{\rm 
trap}(\tau)$ in 
the interval $\tau\in[0,t]$ is less 
than $x$, i.e.,
\begin{equation}
p(t|x)={\rm Prob.}\left[M(t)< x\right].
\label{trap.4}
\end{equation} 
Thus the probability density of the maximum is $\partial_x p(t|x)$
with $x\ge 0$. 
The expected value of the maximum is then given by $E[M(t)]=\int_0^{\infty} 
x\,\partial_x p(t|x)\, dx$. 
Noting that $p(t|x)\to 1$ as $x\to \infty$, it is actually useful to
rewrite this as $E[M(t)]= -\int_0^{\infty} x\, \partial_x[1-p(t|x)]\,dx$
and then do integration by parts giving
\begin{equation}
E[M(t)]= \int_0^{\infty} [1-p(t|x)]\, dx
\label{trap.5}
\end{equation}
Substituting this result in Eq. (\ref{surv.3}) immediately gives the
general result in Eq. (\ref{genrel}).

Let us make a quick remark here.
For a random walk on a $d$-dimensional lattice, starting at the origin, the 
expected number of 
distinct sites visited by a walker up to time $t$ is simply~\cite{Larralde}, 
$\sum_{\vec r} [1-p(t|\vec r)]$, where
$p(t|\vec r)$ is the probability that starting at $\vec r$, the origin
is not visited by the walker up to time $t$. For such a lattice
walk in one dimension, the
expected number of distinct sites visited up to $t$ is thus $\sum_{\vec 
x}[1-p(t|x)]
= 2\,\sum_{x>0}[1-p(t|x)]$. Hence, the quantity $\sum_{x>0}[1-p(t|x)]$ has
the interpretation of half the average number of distinct sites visited by a
walker up to time $t$. The lattice version of the relation (\ref{trap.5}),
where the integral is replaced by a sum over sites on the positive side,
is consistent with this fact since in one dimension the average maximum
up to $t$ is precisely half the average number of distinct sites visited
up to $t$, see \cite{TA, OVKK} and references therein.

\subsection{Expected Maximum $E[M(t)]$ for the CTRW}

In this subsection, we focus on the case when $x_{\rm trap}(\tau)$ is a
CTRW and derive the asymptotic large $t$ expression for $E[M(t)]$ of this
process. The main idea behind this computation is as follows.
Consider first a discrete-time random walk on a continuous line where the position 
$x_n$ of the
walker at step $n$, starting at $x_0=0$, evolves at discrete integer time 
steps 
via the Markov rule 
\begin{equation}
x_n=x_{n-1}+\eta_n
\label{rw.1}
\end{equation}
where $\eta_n$'s are independent and identically distributed (i.i.d) random 
variables, each drawn a a symmetric and continuous probability distribution 
$f(\eta)$. Let $M_n= \max_{0\le k\le n}[\{x_k\}]$ denote the maximum
of this process till step $n$. 
It turns out that an asymptotic expression for the expected maximum $E[M_n]$
for large $n$ can be derived explicitly for arbitrary symmetric and continuous 
jump length distribution $f(\eta)$~\cite{CM}. 
Knowing this explicit result for the discrete-time case, one can then
derive the corresponding expression of $E[M(t)]$ for large $t$ for
a CTRW by using a standard method involving renewal theory.
In fact, it turns out that this renewal method, originally due to
Montroll and Scher~\cite{MS}, is rather general 
in the sense that if one knows explicitly the expression for the
expectation of any observable in the discrete-time jump
process, the corresponding expectation value for the CTRW can be
derived in a straightforward manner, as illustrated below. This
fact sometimes goes by the name `subordination property' 
in the literature~\cite{MK}.  
\begin{figure}
 \includegraphics[width=0.5\textwidth]{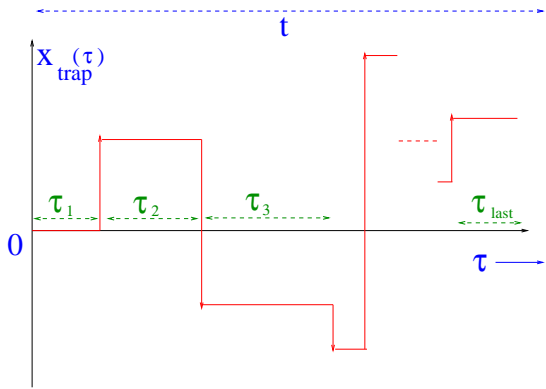}
\caption{ A typical trajectory $x_{\rm trap}(\tau)$ of a CTRW up to a
total time $t$. The process
starts at time $\tau=0$ at the origin and the first interval
$\tau_1$ also starts at $\tau=0$. The trap waits at a 
point in space for a
random time $\tau$ drawn from a distribution $\psi(\tau)$ and then
jumps by a distance $\eta$ drawn from a symmetric and continuous jump
length distribution $f(\eta)$. The successive time intervals $\{\tau_1,
\tau_2,\tau_n, \tau_{\rm last}\}$ spanning the total time $t$ are
statistically independent.} 
\label{ctrw.fig}
\end{figure}

\subsubsection{Subordination property:}

Let $O_n$ be the expected value of any observable associated
with the discrete-time process in Eq. (\ref{rw.1}).
For a corresponding CTRW, these steps do
not occur `at every tick of the clock', but rather at random points
in time separated by random variables $\{\tau_n\}$, called `waiting
times'. The waiting times $\tau_n$'s are i.i.d variables each drawn
from a waiting time density (WTD) denoted by $\psi(\tau)$. 
For a fixed total time $t$, the number of `kicks' $n$ is therefore a random 
variable and let $Q(n|t)$ denote the probability that there are exactly
$n$ `kicks' in time $t$. The continuous-time expectation
value $O(t)$ of the observable in CTRW is then related to its
discrete counterpart $O_n$ by the relation
\begin{equation}
O(t)= \sum_{n=0}^{\infty} O_n\, Q(n|t).
\label{renewal.1}
\end{equation} 

To compute $Q(n|t)$ one can use the renewal property, i.e., the fact that
successive intervals are statistically independent. Consider the
interval $[0,t]$, broken up 
into $(n+1)$ intervals (see Fig. (\ref{ctrw.fig})). The first $n$ intervals are 
denoted by 
$\tau_1$, $\tau_2$, $\dots$, $\tau_n$ and the last interval 
$\tau_{\rm last}=t-\sum_{i=1}^n 
\tau_i$ which is free of `kicks'. While each of $\tau_i$'s (for $i=1,2,\dots,n$)
is drawn from $\psi(\tau)$, the distribution of the last interval
$\tau_{\rm last}$ is slightly different (see Fig. (\ref{ctrw.fig})). 
The probability that an interval $\tau$ is
`kick' free is simply $\phi(\tau)=\int_{\tau}^{\infty} \psi(\tau)\, 
d\tau=1-\int_0^{\tau} \psi(\tau')\, d\tau'$.
Thus, $\tau_{\rm last}$ is drawn from the density function $\phi(\tau)$, while
the preceding ones are each drawn independently from $\psi(\tau)$. Then
the probability $Q(n|t)$ is given by the convolution
\begin{eqnarray}
Q(n|t)= \int_0^\infty\dots\int_0^{\infty} & d\tau_1\,d\tau_2\,\dots
d\tau_n\,  
d\tau_{\rm last}\, \psi(\tau_1)\,\psi(\tau_2)\dots\psi(\tau_n)\,\phi(\tau_{\rm 
last})\nonumber \\
 &\times \delta\left(\tau_1+\tau_2+\dots+\tau_n+\tau_{\rm last}-t\right).
\label{convol.1}
\end{eqnarray} 
Convolutions factorize under a Laplace transform. Denoting the
Laplace transform of the WTD by $\tilde{\psi}(u)$ and noting that
$\int_{0}^{\infty}dt\,e^{-ut}\,\phi(t)=(1-\tilde{\psi}(u))/u$, one gets 
\begin{equation}\label{trans1}
\tilde{O}(u)\equiv
\int_{0}^{\infty}dt\, 
e^{-ut}\, 
O(t)=\frac{1-\tilde{\psi}(u)}{u}\sum_{n=0}^{\infty}
O_n\left[\tilde{\psi}(u)\right]^n.     
\end{equation}
Now since $0\leq\tilde{\psi}<1$ for $u>0$, the sum on the right hand
side of Eq. (\ref{trans1}) can be interpreted as the
generating function
$h(s)\equiv\sum_{n=0}^{\infty}O_n\, s^n$ of $O_n$ with
$s={\tilde{\psi}}(u)$. 
Then the Laplace transform $\tilde{O}(u)$ of the continuous-time quantity $O(t)$
is related to the generating function $h(s)$ of the
discrete-time process via Eq. (\ref{trans1}).

\subsubsection{Typical displacement of a trap:}

As an example of the subordination method, let us consider a discrete-time 
process (\ref{rw.1}) where
the jump length distribution $f(\eta)$ has a power-law tail for large $|\eta|$
\begin{equation}
f(\eta)\sim \frac{A_1}{|\eta|^{1+\nu}}
\label{levy1}
\end{equation}
with $\nu>0$. 
The variance of the jump length distribution $\sigma^2= \int_{-\infty}^{\infty} 
\eta^2\, 
f(\eta)\, d\eta$ is finite if $\nu>2$ and infinite if $\nu<2$.
For $\nu<2$, this process is
the so called L\'evy flights~\cite{Bertoin,BG,MK}.
Let $P_n(x)$ denote the distribution of the position of the trap
after $n$ steps, starting at the origin. It is then well known
and is easy to show that $P_n(x)$, for large $n$, has the scaling form~\cite{BG}
\begin{equation}
P_n(x)= \frac{1}{n^{1/\mu}}\, {\cal L}_{\mu}\left(\frac{x}{n^{1/\mu}}\right)
\label{levyscaling.1}
\end{equation}
where the scaling function, for large $|z|$, decays as  ${\cal L}_{\mu}(z)\sim 
|z|^{-1-\mu}$ 
for $0<\mu<2$ and is Gaussian for $\mu\ge 2$. Thus, the typical displacement
of the trap after step $n$ scales, for large $n$, as
\begin{eqnarray}
x_{\rm typ}(n) & \sim &  n^{1/\nu} \quad {\rm for}\quad 0<\nu\le 
2\,\,(\rm 
superdiffusive) \label{superdiff.1}\\
&\sim &  n^{1/2} \quad {\rm for}\quad \nu\ge 2\,\,(\rm    
diffusive) \label{diff.1}
\end{eqnarray}
Let us also consider a WTD $\psi(\tau)$ that has a power-law tail for large 
$\tau$
\begin{equation}
\psi(\tau)\to \frac{A}{\tau^{1+\alpha}}
\label{WTDpower}
\end{equation}
with $\alpha>0$. The mean waiting time $\mu=\int_{0}^{\infty}\tau\, \psi(\tau)\, 
d\tau$ is finite if $\alpha>1$ and infinite if $\alpha<1$. 
Correspondingly, the leading asymptotic  
behavior of the Laplace
transform ${\tilde \psi}(u)$ of the WTD can be computed near $u\to 0$ (see,   
e.g., appendix 2 of Ref. \cite{EMZ}) as
\begin{equation}\label{wtd_lt}
\tilde{\psi}(u)\approx \left\{\begin{array}{rl}1-\mu\, u  & \textrm{for finite }
    \mu \\
    \vspace{1.5mm}\\
  1-A\, |\Gamma(-\alpha)|u^{\alpha} & \textrm{for infinite } \mu 
\end{array}\right.
\end{equation}
Replacing $O_n$ by $x_{\rm typ}(n)$ in (\ref{trans1})
and using the 
above asymptotic expressions one can easily check that $x_{\rm typ}(t)$, the
typical displacement of the trap undergoing CTRW,  
behaves asymptotically in different ways 
depending on the following four cases
\begin{enumerate}
  \item both $\sigma$ and $\mu$ finite
  \item $\sigma$ finite, $\mu$ infinite
  \item $\sigma$ infinite but $\mu$ finite 
  \item both $\sigma$ and $\mu$ infinite
\end{enumerate}
which, for power-law jump length distribution $f(\eta)$ in Eq. (\ref{levy1}) and 
power-law WTD in Eq. (\ref{WTDpower}) correspond to
\begin{enumerate}
  \item $\nu>2$ and $\alpha>1$
  \item $\nu>2$ and $\alpha<1$
  \item $\nu\le 2$ and $\alpha>1$
  \item $\nu\le 2$ and $\alpha<1$.
\end{enumerate}
This gives
\begin{equation}\label{rms_four_cases}
x_{\rm typ}(t)\sim \left\{\begin{array}{rl} 
t^{1/2} & \mathrm{case\ 1}\\ \vspace*{1pt}\\
t^{\alpha/2} & \mathrm{case\ 2}\\ \vspace*{1pt}\\
t^{1/\nu} & \mathrm{case\ 3}\\ \vspace*{1pt}\\
t^{\alpha/\nu} & \mathrm{case\ 4}
 \end{array}\right.
\end{equation}
where the cases refer to the four different situations mentioned above.
Note, in particular, that by appropriately 
choosing the input exponents $\alpha$ and $\nu$, one can generate
diffusive (case 1), subdiffusive (case 2), superdiffusive (case 3),
as well as a completely anomalous (case 4) where one can get both
subdiffusive and  superdiffusive behavior by appropriately choosing
$\alpha/\nu$.  

As an aside, another recent example where this subordination property has
been used concerns the statistics of records. For a discrete-time jump process
in Eq. (\ref{rw.1}), the statistics of the number of records
up to step $n$ turn out to be completely universal, i.e.,  
independent of the jump length distribution $f(\eta)$ as long it 
is symmetric and continuous~\cite{MZrecord}. For instance,
the average number of records up to step $n$, grows as
$\langle R_n\rangle \approx \sqrt{4n/\pi}$ for large $n$~\cite{MZrecord}.
Using the subordination property mentioned above and by choosing
$O_n=R_n$, the average number of records up to time $t$ for a CTRW
with power-law WTD as in Eq. (\ref{WTDpower}) has been computed 
recently~\cite{Sabha}. 

\subsubsection {Expected maximum of the trap motion:}

Having obtained the late time scaling of the typical displacement
of a trap undergoing CTRW, we now turn to computing the expected maximum
$E[M(t)]$ of the trap motion $x_{\rm trap}(t)$. In order to use
the general subordination technique mentioned above, we then need
to know the behavior of $E[M_n]$ for large $n$ of the discrete-time process 
(\ref{rw.1}) where $M_n$ denote the maximum up to step $n$.
Fortunately this can be done as illustrated below.

Computing explicitly the full distribution of $M_n$ for arbitrary $f(\eta)$
is, in general, a hard problem~\cite{CM,SM09}. However, there exists
a general formula known as Pollaczek-Spitzer formula~\cite{P,S} that 
reads~\cite{CM}
\begin{equation} 
\sum_{n=0}^{\infty} s^n\, E[e^{-\rho M_n}]= 
\frac{1}{\sqrt{1-s}}\,\phi(s,\rho) 
\label{ps1}
\end{equation}
where
\begin{equation} 
\phi(s,\rho)=\exp\left[-\frac{\rho}{\pi}\,\int_0^{\infty} \frac{\ln
    \left(1-s {\hat f}(k)\right)}{\rho^2+k^2}\, dk\right],
\label{ps1b}
\end{equation}
and ${\hat f}(k)= \int_{-\infty}^{\infty} f(\eta)\, e^{i\,k\,\eta}\, d\eta$
is the Fourier transform of $f(\eta)$. From the expression (\ref{ps1}), the
generating function of the expected maximum can, in principle, be obtained
via the formula
\begin{equation}
h(s)= \sum_{n=0}^{\infty} s^n\, E[M_n]= -\frac{1}{\sqrt{1-s}}\,\frac{\partial 
\phi(s,\rho)}{\partial \rho}\Big|_{\rho=0}.
\label{ps2}
\end{equation}
Replacing $O_n$ by $E[M_n]$ in the general formula (\ref{trans1}), we then
have for CTRW
\begin{equation}
\int_0^{\infty} dt\, e^{-u\,t}\, E[M(t)]= \frac{1-\tilde{\psi}(u)}{u}\, 
h\left({\tilde \psi}(u)\right).
\label{lt1}
\end{equation}

To extract the leading asymptotic behavior for large $t$, one needs to
analyze the right hand side of Eq. (\ref{lt1}) in the limit $u\to 0$, or 
equivalently $h(s)$ in Eq. (\ref{ps2}) in the limit $s\to 1$.  
Extracting the leading singularity of the right hand side of (\ref{ps2}) in the
limit $s\to 1$ turns out to be rather nontrivial~\cite{CM}. 
For jump length distributions
with finite second moment $\sigma^2=\int_{-\infty}^{\infty} \eta^2\, 
f(\eta)\,d\eta$, the principal singular behavior of $h(s)$ near 
$s=1$ can be extracted explicitly~\cite{CM}

\begin{equation}\label{finites_zt}
 h(s)=\frac{\sigma}{\sqrt{2}}\frac{1}{(1-s)^{3/2}}+\frac{1}{\pi(1-s)}\int\frac{dk}{k^2}\ln\left[\frac{2}{\sigma^2}\left(\frac{1-\hat{f}(k)}{k^2}\right)\right]  
 + \mathcal{O}\left(\frac{1}{\sqrt{1-s}}\right)
\end{equation} 

In case when $\sigma^2$ is infinite, as for instance the case for L\'evy 
flights in Eq. (\ref{levy1}) with $\nu<2$, where
$\hat{f}(k)\approx 1- |ak|^\nu$ to
leading order for $k\to 0$, the asymptotic behavior of
$h(s)$ near $s=1$ turns out to be different from Eq. (\ref{finites_zt}). 
For example, for $1<\nu \le 2$, one can show~\cite{CM}
\begin{equation}\label{infinites_zt}
h(s)=\frac{aB(1/\nu,
  1-1/\nu)}{\pi(1-s)^{1+1/\nu}}+\frac{1}{\pi(1-s)}\int_{0}^{\infty}\frac{dk}{k^2}\ln\left(\frac{1-s\hat{f}(k)}{(ak)^{\nu}}\right) 
+ \mathcal{O}\left(\frac{1}{(1-s)^{1/\nu}}\right)
\end{equation}
where $B(a,b)=\Gamma[a]\Gamma[b]/\Gamma[a+b]$ is the standard Beta function.

For the case $0<\nu\le 1$, it turns out $E[M_n]$ is infinite for any $n>0$. 
Subsequently $E[M(t)]$ is also infinite for any finite $t$. Consequently, for 
$0<\nu\le 1$, the survival probability $P_s(t)$, via
the relation (\ref{genrel}), is $0$ for any finite time $t$ (the target
is killed instantaneously). This rather pathological case will not
be discussed further. 

\begin{figure}
 \includegraphics[width=0.5\textwidth]{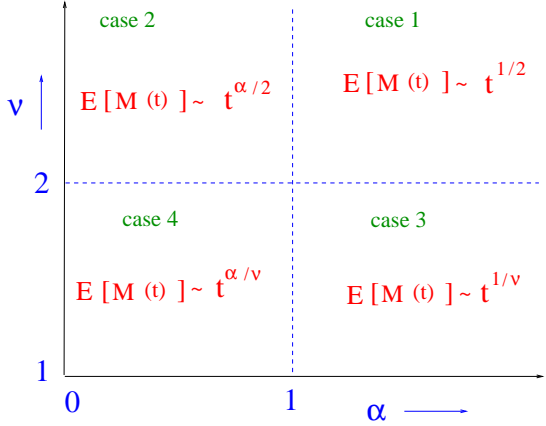}
\caption{The four different asymptotic power-law growth of $E[M(t)]$ for 
large $t$ in the four cases in the $(\alpha-\nu)$ plane is summarized.}
\label{phasedia.fig}
\end{figure}

We now need to substitute these asymptotic $s\to 1$ behaviors (for the two cases
in Eqs. (\ref{finites_zt}) and (\ref{infinites_zt})) in Eq. (\ref{lt1})
and analyse the large $t$ behavior of $E[M(t)]$. To carry out
the asymptotic analysis, 
one again needs to distinguish four different cases mentioned in the
previous subsection, namely (1) both $\sigma$ and $\mu$ finite (2)
$\sigma$ finite but $\mu$ infinite (3) $\sigma$ infinite but $\mu$ finite
(4) both $\sigma$ and $\mu$ infinite, where $\sigma^2$ and $\mu$ refer
respectively to the second moment of the jump length distribution $f(\eta)$ and
the first moment of the WTD $\psi(\tau)$.
When both $f(\eta)$ and $\psi(\tau)$
have power-law tails as in Eqs. (\ref{levy1}) and (\ref{WTDpower}) respectively,
the above $4$ cases correspond to $4$ regions in the $(\alpha-\nu)$ plane
(1) $\nu>2$ and $\alpha>1$ (2) $\nu>2$, $\alpha<1$ (3) $\nu\le 2$, $\alpha>1$  
and (4) $\nu\le 2$, $\alpha<1$. 
As mentioned before, we will not discuss the pathological case $0<\nu\le 1$
and will restrict ourselves in the $(\alpha-\nu)$ plane only for $\nu>1$ and
$\alpha>0$ (see Fig. (\ref{phasedia.fig})).

\begin{figure}
  \includegraphics[width=0.46\textwidth]{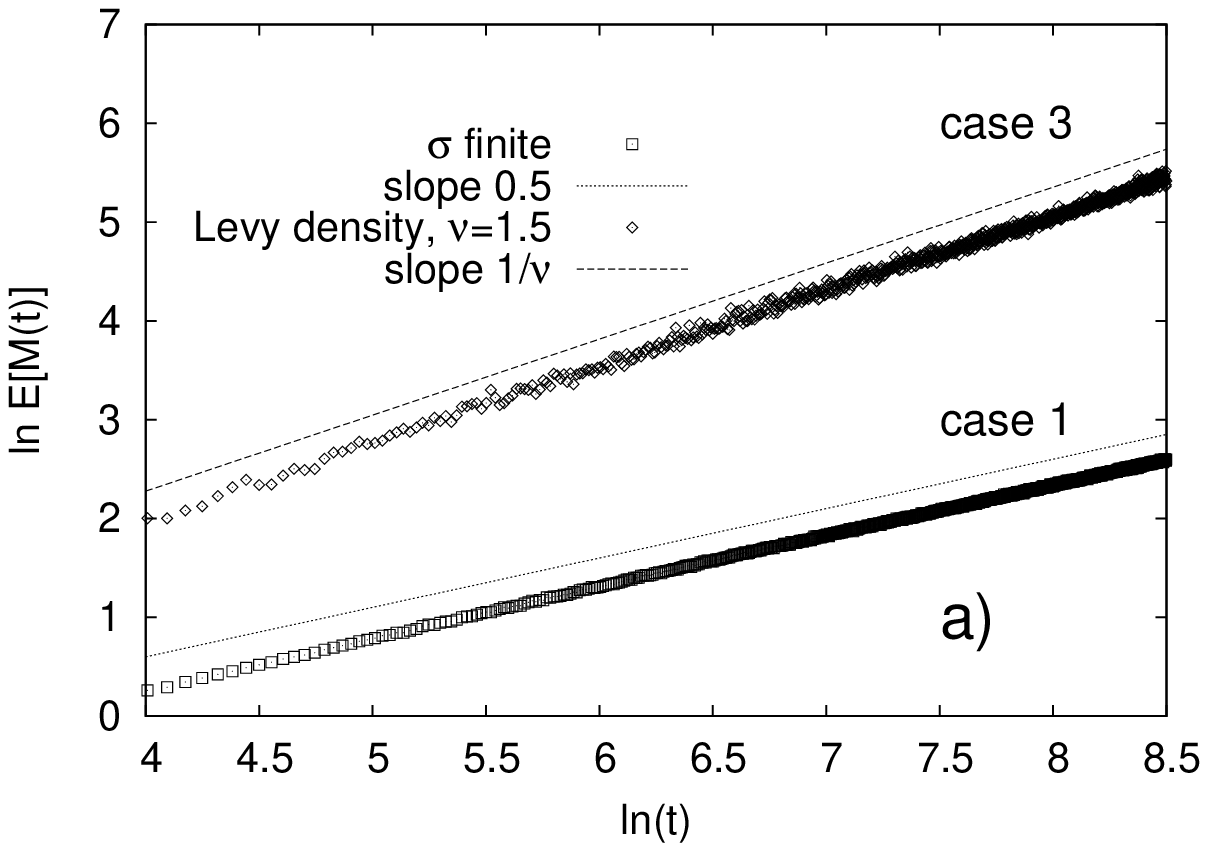}
  \hfill
  \includegraphics[width=0.46\textwidth]{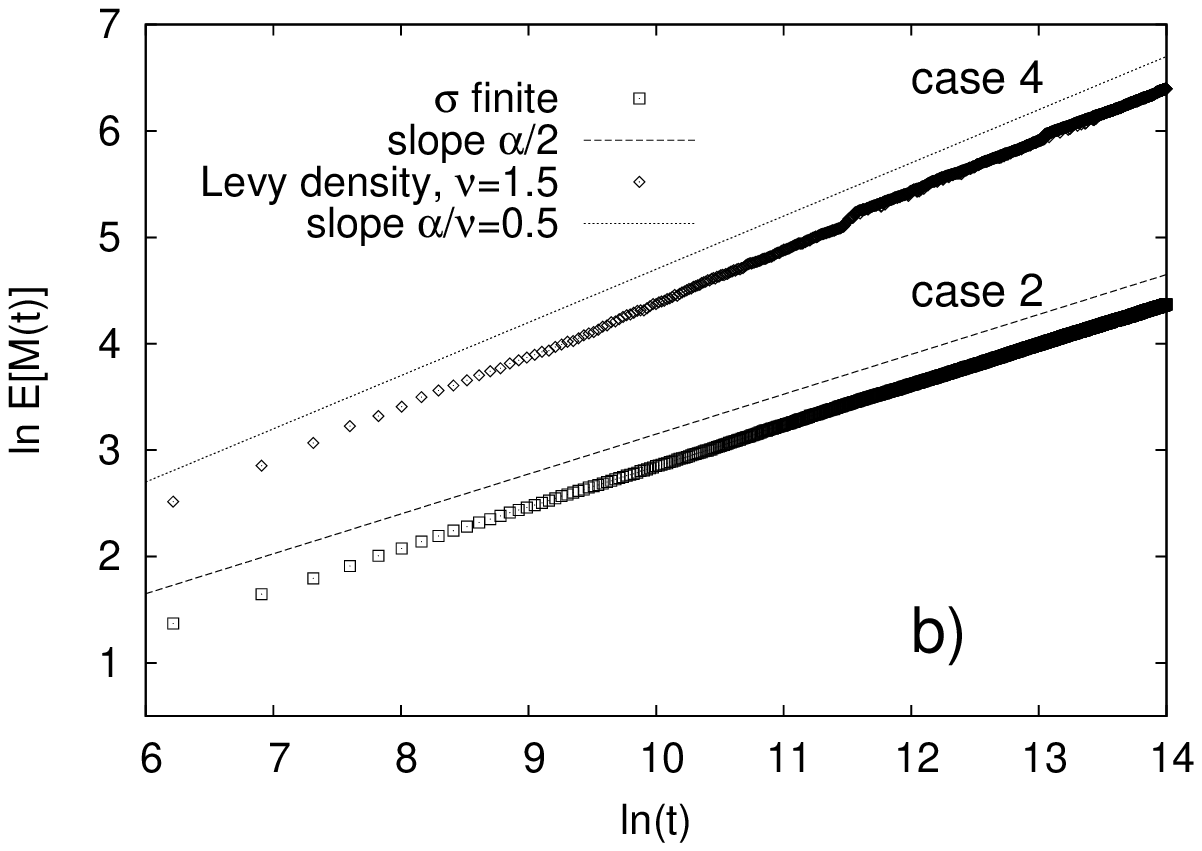}
  \caption{\label{scaling} Expected maximum of a CTRW as function of
    $t$. On the log-log scale, the exponent of $t$ can be read off from
    the slope. Fig. (4a) corresponds to the case of finite $\mu$ 
while
Fig. (4b) corresponds to infinite $\mu$. The bottom curve of 
Fig. (4a)
corresponds to case 1 where both $\mu$ and $\sigma$ are finite and the
exponent is $1/2$ (see Eq. (\ref{four_cases})). The top curve of Fig. 
(4a)
corresponds to case 3 where $\mu$ is finite but $\sigma$ is infinite--
the exponent is $1/\nu$. Similarly, in Fig. (4b) the bottom 
curve
corresponds to case 2 (where $\mu$ is infinite but $\sigma$ is finite
and the exponent is $\alpha/2$ as in Eq. (\ref{four_cases})) while
the top curve corresponds to case 4 (where both $\mu$ and $\sigma$ are infinite
and the exponent is $\alpha/\nu$). Note, in particular, that
in the bottom curve of Fig. (4b), by choosing $\alpha/\nu=1/2$
one obtains the diffusive behavior even though the process is totally 
anomalous. Simulations used a
    L\'evy-stable or standard normal density for the jump lengths and
    either exponential or Pareto distributed waiting
    times. Parameters are given in the figure.}
\end{figure}

The next step is to plug the
appropriate asymptotic expression for the Laplace transform of the
 WTD from Eq. (\ref{wtd_lt}) into the corresponding expression for the
 generating function of $E(M_n)$ in Eqs. (\ref{finites_zt}) and
 (\ref{infinites_zt}) and subsequently invert the Laplace transform
 in the asymptotic large $t$ limit. One needs to use
 the Tauberian theorem, cf. \cite{Feller,
 Hughes} that states that a Laplace space behavior of $u^{-\rho}$ for
 $u \ll 1$ corresponds to a $t^{\rho-1}$ behavior for $t \gg 1$ under
 inversion of the Laplace transform. It turns out that three constants
appear in the final expressions for $E[M(t)]$ which we denote below
as 

\begin{eqnarray}
  K\equiv &\frac{1}{\pi}\int_{0}^{\infty}\frac{dk}{k^2}\ln
\left[\frac{2}{\sigma^2}\left(\frac{1-\hat{f}(k)}{k^2}\right)\right],
\label{amplitudeK}\\
  I\equiv &
  \frac{1}{\pi}\int_{0}^{\infty}\frac{dk}{k^2}\ln\left(\frac{1-\hat{f}(k)}{(ak)^\nu}\right) 
  \label{amplitudeI}\\
J\equiv &\frac{aB(1/\nu, 1-1/\nu)}{\pi\Gamma(1+\alpha/\nu)
\left(A|\Gamma(-\alpha)|\right)^{\nu}}\label{amplitudeJ}
\end{eqnarray}
with $B(a,b)$ being the standard Beta function. In the definition of Eq. 
(\ref{amplitudeK})
we assume $\nu>2$ ($\sigma^2$ finite) and in Eq.  
(\ref{amplitudeI}), one assumes $1<\nu<2$.
We then obtain for large $t$ the following asymptotic expressions for
$E[M(t)]$ in the four cases  

\begin{equation}\label{four_cases}
  E[M(t)]\approx \left\{\begin{array}{rl} \frac{\sqrt{2}\sigma}{\sqrt{\pi
        \mu}}\,t^{1/2} +K & \mathrm{case\ 1}\\ \vspace*{1pt}\\
    \frac{\sigma}{\Gamma(1+\alpha/2)\sqrt{2A|\Gamma(-\alpha)|}}\,t^{\alpha/2}
    +K & \mathrm{case\ 2}\\ \vspace*{1pt}\\
    \frac{a\nu\Gamma(1-1/\nu)}{\pi(\mu)^{1/\nu}}\,t^{1/\nu} +I & \mathrm{case\ 3}
    \\ \vspace*{1pt}\\
    J\,t^{\alpha/\nu}+I & \mathrm{case\ 4}
  \end{array} \right.
\end{equation} 
where the cases refer to the four different situations mentioned above.
Note that the result in case 1 also holds for arbitrary distributions $f(\eta)$ and 
$\psi(\tau)$ such
that $\sigma^2$ and $\mu$ are finite.
These $4$ different asymptotic power-laws for the growth of $E[M(t)]$ in the
$(\alpha-\nu)$ plane are
summarized in Fig. (\ref{phasedia.fig}).
Note that the leading asymptotic growth of $E[M(t)]$ for large $t$ is
similar to that of the typical displacement in Eq. (\ref{rms_four_cases})
in all the $4$ regions, except the prefactors of the power-law growth
of $E[M(t)]$ and $x_{\rm typ}(t)$ are different.   
As in the typical case, $E[M(t)]$ shows the usual diffusive scaling to 
leading order in case 1, subdiffusive in case 2, superdiffusive in
case 3, while in case 4 it scales like
$t^{\alpha/\nu}$. Interestingly, this totally anomalous case 4 of
L\'evy flights interspersed with long waiting times can exhibit
diffusive scaling, for any choice of $\alpha$ and $1<\nu\le 2$ but
with $\alpha/\nu=1/2$.
We have verified the analytical predictions in Eq. (\ref{four_cases})
via numerical simulations.
The results are shown in fig
\ref{scaling} where different types of anomalous scaling behaviors are
compared. Numerical results are in excellent agreement with analytical 
predictions.

\section{Asymptotic Survival Probability: Analytical Predictions and
  Numerical  Simulations}

Plugging the expressions for $E[M(t)]$ for the four cases considered
above in the general relation in Eq. (\ref{genrel}), we obtain the
following large $t$ behavior for the average survival probability $P_s(t)$

\begin{equation}\label{surv_prob}
  P_s(t)\approx \left\{\begin{array}{rl}\exp\left[-2\rho \left(\frac{\sqrt{2}\sigma}{\sqrt{\pi
        \mu}}t^{1/2} +K\right)\right] & \mathrm{case\ 1} \\ \vspace*{1pt} \\
\exp\left[-2\rho
  \left(\frac{\sigma}{\Gamma(1+\alpha/2)\sqrt{2A|\Gamma(-\alpha)|}}t^{\alpha/2}
    +K\right)\right] & \mathrm{case\ 2} \\ \vspace{1pt} \\
\exp\left[-2\rho
  \left(\frac{a\nu\Gamma(1-1/\nu)}{\pi(\mu)^{1/\nu}}t^{1/\nu}
    +I\right)\right] & \mathrm{case\ 3}\\ \vspace*{1pt} \\
\exp\left[-2\rho \left(Jt^{\alpha/\nu}+I\right)\right] & \mathrm{case\
4} 
\end{array}\right.
\end{equation}
where the constants $K$, $I$ and $J$ are given respectively in Eqs. 
(\ref{amplitudeK}), (\ref{amplitudeJ}) and (\ref{amplitudeJ}).
Thus, in all four cases, one finds generically the stretched exponential
behavior for large $t$
\begin{equation}
P_s(t) \approx a\, \exp\left[-b\, t^{\theta}\right] 
\label{ste.1}
\end{equation}
except that the amplitude $a$, the constant $b$ and the stretching exponent
$\theta$ differs in the $4$ cases. In the $4$ different cases, the exponent 
$\theta$ is respectively (1) $\theta=1/2$ (2) $ \theta=\alpha/2$ with $\alpha<1$
(3) $\theta=1/\nu$ with $1<\nu\le 2$ and (4) $\theta=\alpha/\nu$ with
$\alpha<1$ and $1<\nu\le 2$. We also emphasize that our method
yields an exact result even for the amplitude $a$ in front of the stretched
exponential in Eq. (\ref{ste.1}). 
Usually this is much harder to obtain as
it corresponds to computing the subleading correction term in $E[M(t)]$.
In the four respective cases, we get the amplitudes 
\begin{enumerate}
\item  $a=\exp[-2\,\rho\, K]$
\item  $a= \exp[-2\,\rho\, K]$ 
\item  $a= \exp[-2\,\rho\, I]$
\item  $a= \exp[-2\,\rho\,I]$
\end{enumerate}
where the constants $K$ and $I$ are given respectively in Eqs. (\ref{amplitudeK})
and (\ref{amplitudeI}).

\subsection{Recovering known results for diffusive and subdiffusive
  traps} 

From our general result in Eq. (\ref{surv_prob}), it is easy to recover
the known results for the Brownian (diffusive)  and subdiffusive traps first 
derived in \cite{TA} and \cite{YL} respectively. Consider first the
Brownian case. 
To recover this result corresponding to Brownian motion of a trap, we need
to focus on case 1 of our result in Eq. (\ref{surv_prob}) and take
the Brownian scaling limit where both mean square jump length
$\sigma^2$ and mean  
waiting time $\mu$ tend to zero, while their ratio
$D= \sigma^2/2\mu$ stays constant. In order to take this limit cleanly, it is 
convenient to first make a change of variable
$k\to\sigma k/\sqrt{2}$ in the expression for the constant $K$ in
(\ref{amplitudeK}) to get
\begin{equation}\label{another_K}
K=\frac{\sigma}{\sqrt{2}\pi}\int_{0}^{\infty}\frac{dk}{k^2}
\ln\left(\frac{1-{\hat f}(\sqrt{2}k/\sigma)}{k^2}\right). 
\end{equation}
Note that $K$ is independent of $\mu$.
Since ${\hat f}(k)\to 0$ as $k\to \infty$, we can safely take the
limit $\sigma\to 0$ in Eq. (\ref{another_K}) and obtain $K=0$.
From case 1 of Eq. (\ref{surv_prob}), we then recover the
Brownian result: $P_s(t)\to\exp (-2\rho\sqrt{4Dt/\pi})$.

To recover the result for the continuous-time subdiffusive case in \cite{YL},
we need to take the scaling limit of our result in Eq. (\ref{surv_prob})
corresponding to case 2. As in the Brownian case, we need to take
the $\sigma^2\to 0$ limit. However, unlike in the Brownian case,
here the
mean waiting time $\mu$ is already infinite. So, the only way a sensible limit
of line 2 in Eq. (\ref{surv_prob}) can be reached if $2 A |\Gamma(-\alpha)|$
goes to zero while $\sigma^2\to 0$, but their ratio
$D_{\alpha}\equiv \sigma^2/(2A|\Gamma(-\alpha)|)$ remains fixed.
This ratio $D_{\alpha}$ can be called the generalized diffusion constant
since it can be shown that it appears also in the amplitude of the mean-square 
displacement
in case 2 (with $\alpha<1$)
\begin{equation}
\langle x_{\rm trap}^2(t)\rangle \approx \frac{2 D_{\alpha}}{\Gamma(1+\alpha)}\, 
t^{\alpha}\, .
\label{gendiff}
\end{equation}
This relation can be thought of as the generalization of the standard
diffusion relation
$\langle x_{\rm trap}^2(t)\rangle = 2 D t$ with
$D$ being the standard
diffusion constant. Indeed Eq. (\ref{gendiff}) reduces to the standard
diffusion relation when $\alpha\to 1$.
Taking this scaling limit in line 2 of Eq. (\ref{surv_prob}), we thus recover
the result of Yuste and Lindenberg~\cite{YL}

\begin{equation}
  P_s(t)\to
  \exp\left[-\frac{2\rho\sqrt{D_{\alpha}}}{\Gamma(1+\alpha/2)}t^{\alpha/2}\right]. 
\end{equation} 

Our general result in Eq. (\ref{surv_prob}), apart from recovering
known results in the appropriate limits as shown above, also provides
new results for the superdiffusive case (which can be achieved in
cases 3 and 4). 
In addition, our result also provides not just the leading asymptotic behaviors
in all cases, 
but 
also the subleading order corrections in the form of non-universal
amplitudes. 

\subsection{Numerical simulations}

\begin{figure}
  \includegraphics[width=0.46\textwidth]{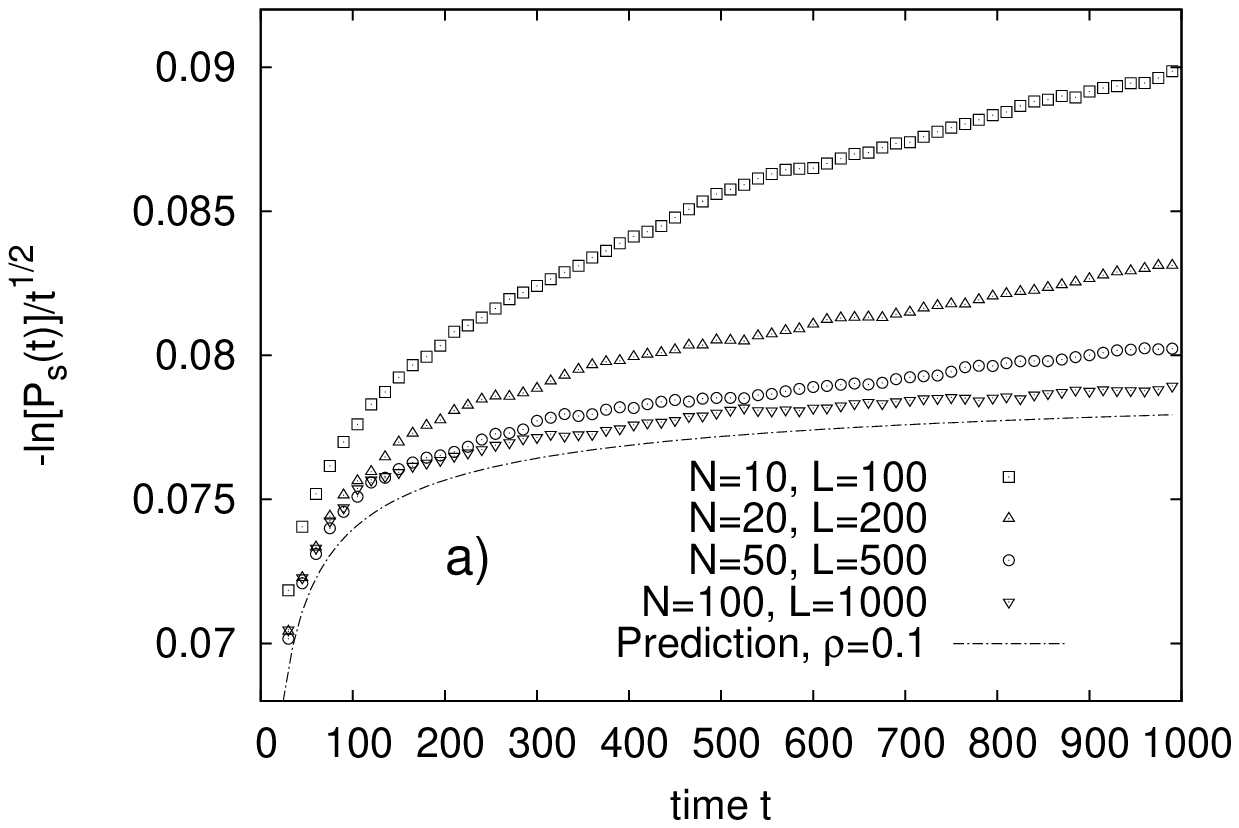}\hfill
  \includegraphics[width=0.48\textwidth]{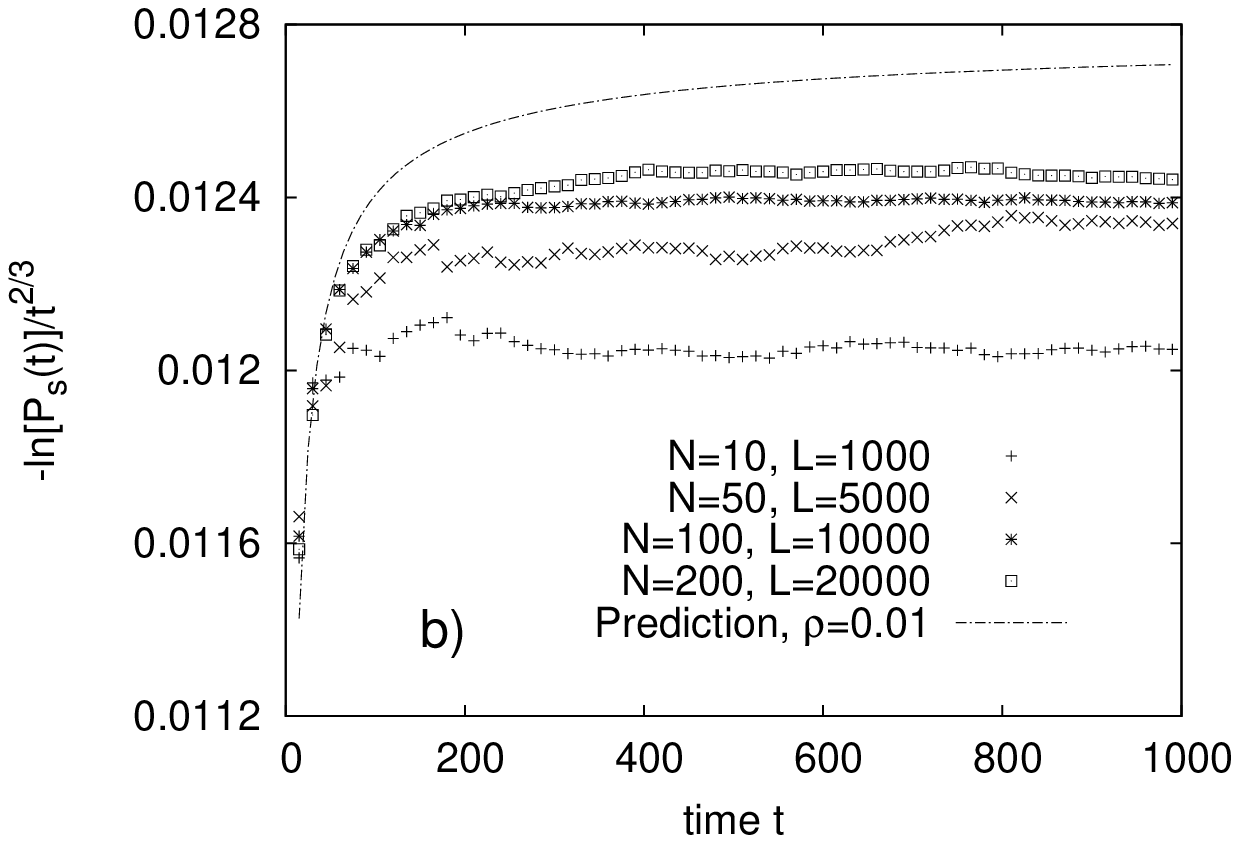}
  \caption{\label{num_check} \textbf{Left:} For standard normal
    distributed jumps length with $\sigma^2=1$. This corresponds to case 1.
The dotted line shows the
analytical behavior 
\textbf{Right:} For L\'evy jump length distribution with Fourier transform
${\hat f}(k)= \exp[- (a|k|)^{\nu}]$ with the choice
 $\nu=1.5$ and $a=1$. While for the
    Gaussian case, the survival probability was averaged over $10^5$
    realizations, for the L\'evy-distribution, between $10^6$ ($N=10, L=1000$) and
    $10^7$ (all others) realizations were used, since due to the possibly large
    jumps, only very little realizations survive up to long times. For
    the same reason, the density of traps is a factor of $10$ lower
    than for the Gaussian case. }
\end{figure}

To confirm our analytical results, we also performed numerical simulations.
Due to the computational effort of simulating large
systems, only a finite number of $N$ particles uniformly distributed
over the interval $(0,L)$ with $N/L=\rho$ fixed were
considered. The reason for only considering traps on one side of the
origin in just computational convenience. For better accuracy of
comparison, fig \ref{num_check} compares $-\ln (P_s(t))/t^{\theta}$ to
numerical simulations with the exponent $\theta$ taking the value
corresponding to 
the case considered, see Eq. (\ref{surv_prob}). In all cases,
$-\ln (P_s(t))/t^{\theta}$ is of the form $\rho a_1 +\rho a_2/t^{\theta}$
with constants $a_1$ and $a_2$ depending on the cases. 
The missing prefactor $2$ infront of the trap density $\rho$ is due to
the asymmetry of having only traps on one side of the origin.
 
Simulations were then performed for increasing values of
$N$ and $L$ to show that as the effects of finite system size become
less important, the simulated curves converge onto the analytical
predictions, see Fig. \ref{num_check}. The simulations were performed
with the WTD $\psi(\tau)=\delta(\tau-1)$, thus leaving us with only
cases 1 and 3 to distinguish. The influence of infinite mean waiting
time on the expected value of the maximum of a CTRW has already been
illustrated numerically in Fig. \ref{scaling}, so it suffices to
study the survival probability of a target surrounded by traps
performing a random walk or L\'evy flight with finite $\mu$. 
From Fig. \ref{num_check}, one sees that with increasing size $L$, the
numerical curves approach the analytically predicted behaviors shown by
dotted lines. 

\section{Conclusion}

To summarize, we have studied analytically, in one dimension, the survival 
probability $P_s(t)$
of an immobile target at the origin surrounded by a sea of independent
mobile traps  
initially distributed uniformly with density $\rho$ on the line. We have
shown that for a rather general stochastic motion $x_{\rm trap}(t)$ of
each trap, there is a very simple general relation, $P_s(t)= \exp\left[-2\rho\, 
E[M(t)]\right]$, that relates the survival probability to the expected
maximum of the process $x_{\rm trap}(t)$. This general result recovers all
earlier known special cases of $x_{\rm trap}(t)$. In addition, we
have presented new result when $x_{\rm trap}(t)$ is a continuous-time
random walk (CTRW) with arbitrary jump length distribution $f(\eta)$ and 
with arbitrary
waiting time distribution $\psi(\tau)$.  
By choosing these two input distributions of CTRW one can generate a variety
of stochastic motions of the trap, both subdiffusive and superdiffusive.
For such a general CTRW,
we were able to evaluate exactly the two leading terms in the
asymptotic behavior of $E[M(t)]$ for large $t$ and plugging
this in our general relation, we were able to compute the
two leading terms for the large $t$ behavior of $P_s(t)$. 
Generically, we found that for large $t$
\begin{equation}
P_s(t)\sim a\, \exp[-b\, t^{\theta}]
\label{genresult}
\end{equation}
where the exponent $\theta$, as well as the two constants $a$ and $b$
are computed exactly in this paper when $x_{\rm trap}(t)$ is
a CTRW with arbitrary $f(\eta)$ and $\psi(\tau)$. 
Previous results in the literature were known only for diffusive and 
subdiffusive $x_{\rm trap}(t)$ (case 1 and case 2 in Fig. (\ref{phasedia.fig})). 
Our result is more general and in particular,
the exact asymptotic expressions
for the two remaining cases (the superdiffusive (case 3) and
totally anomalous (case 4))
are, to our knowledge, new.
In addition, our method allows us to compute
even the nonuniversal amplitude $a$ (and not just the stretching exponent 
$\theta$) exactly in all cases.

There are several future directions in which it may be possible 
to extend our results.
First, it would be interesting to generalize our result
to higher dimensions. The main difficulty in higher dimensions is
that it is difficult to obtain an explicit expression for 
the probability that a discrete-time random walker does not 
hit a static ball (target) of finite radius up to time
$t$~\cite{Ziff1,MCZ}, though 
it may be possible to obtain some exact results in the special case
of $d=3$~\cite{MCZ,ZMC1,ZMC2}. It would be interesting to derive
the result for $P_s(t)$ in this more realistic $d=3$ case.

Secondly, it would be interesting to derive the corresponding result
for $P_s(t)$, 
in one or higher dimensions, in the case when the target itself is 
mobile, e.g., when the target also performs a general CTRW as the trap.
As mentioned in the introduction,
in special cases of diffusive and subdiffusive motion of the target it
is known that the leading asymptotic behavior (but not the subleading)
of $P_s(t)$ does not change from the situation when the target is static.
Is this fact still true when the target performs a general CTRW? It would be 
interesting to prove or disprove this fact.

Finally, it would be interesting to find suitable applications
of our results. In fact, the totally anomalous case 4 (see Fig. 
(\ref{phasedia.fig})) for which we present new exact results,  
may
at first glance seem to be only of academic interest, but has 
recently been realized in a more physical context, namely, in a model
for spread  
of epidemics by human travel \cite{DS}. The
results presented above thus have a natural interpretation as the
probability of staying on the `safe' side of the propagating front
of a disease up to time $t$ and it would be interesting to explore
in detail this interesting possible application in future. 

\vskip 0.3cm

\noindent\textbf{Acknowledgements:} JF thanks the LPTMS for hospitality during
his Diploma thesis in 2007 and acknowledges financial support from
Studienstiftung des deutschen Volkes. S.N.M. acknowledges 
support by ANR grant 2011-BS04-013-01 WALKMAT and by the Indo-French
Centre for the Promotion of Advanced Research under Project 4604-3.

\end{document}